\documentclass[twocolumn,10pt,prl,showpacs]{revtex4-1}

\usepackage{hyperref}
\usepackage{epsfig}
\usepackage{amsmath,amssymb}
\usepackage{verbatim}

\newcommand{\beq}{\begin{equation}}
\newcommand{\eeq}{\end{equation}}
\def\bea#1\eea{\begin{align}#1\end{align}}
\newcommand{\nn}{\nonumber}

\renewcommand{\d}{\textrm{d}}
\newcommand{\w}{\wedge}

\def\del {\partial}

\begin{document}

\title {\Large\bf Minkowski flux vacua of type II supergravities}
 \author{\bf David Andriot$^{a, b}$, Johan Bl{\aa }b\"ack$^{c}$, Thomas Van Riet$^{d}$}
\affiliation{$^{a}$Max-Planck-Institut f\"ur Gravitationsphysik, Albert-Einstein-Institut,\\Am M\"uhlenberg 1, 14467 Potsdam-Golm, Germany}
\affiliation{$^{b}$Institut f\"ur Mathematik, Humboldt-Universit\"at zu Berlin, IRIS-Adlershof,\\Zum Gro\ss en Windkanal 6, 12489 Berlin, Germany}
\affiliation{$^{c}$Institut de Physique Th\'eorique, Universit\'e Paris Saclay,\\ CEA, CNRS, F-91191 Gif sur Yvette, France}
\affiliation{$^{d}$Instituut voor Theoretische Fysica, K.U.Leuven,\\Celestijnenlaan 200D, B-3001 Leuven, Belgium}
\affiliation{\upshape\ttfamily david.andriot@aei.mpg.de, johan.blaback@cea.fr, thomasvr@itf.fys.kuleuven.be}

\begin{abstract}
We study flux compactifications of 10d type II supergravities to 4d Minkowski space-time, supported by parallel orientifold $O_p$-planes with $3\leq p \leq 8$. With some restrictions, the 4d Ricci scalar can be written as a negative sum of squares involving BPS-like conditions. Setting all squares to zero provides automatically a solution to 10d equations of motion. This way, we characterize a broad class, if not the complete set, of Minkowski flux vacua with parallel orientifolds. We conjecture an extension with non-geometric fluxes. None of our results rely on supersymmetry.
\end{abstract}

\pacs{11.25.Mj, 11.25.Wx, 04.65.+e, 04.20.Ex}

\maketitle

\section{I. Context and results}
We consider vacua of ten-dimensional (10d) type II supergravities on a 4d maximally symmetric space-time times a 6d compact manifold, with fluxes. Such vacua are a major framework for string phenomenology, given the role of fluxes in moduli stabilization. Having a complete classification of these vacua would thus be a significant achievement. We make here an essential step towards this goal: with few requirements, we reveal and characterize a broad class of Minkowski flux vacua.

The best understood flux compactifications of type II supergravities are those with just fluxes and no sources. Well-known no-go theorems \cite{Maldacena:2000mw} imply that such reductions only lead to anti-de Sitter (AdS) vacua. Even more, those AdS vacua are no genuine compactifications since they lack a tunable hierarchy between the AdS length scale and the Kaluza-Klein scale \cite{Gautason:2015tig}. Space-time filling orientifold planes are a natural way out of this problem and allow one to find Minkowski vacua \cite{Dasgupta:1999ss, Giddings:2001yu} (see \cite{Grana:2005jc} for a review) or AdS ones with scale separation (see e.g.~\cite{DeWolfe:2005uu}). However such reductions come with an extra level of complication, the backreaction of the orientifolds themselves. In the case of non-intersecting $O_p$-planes (and possibly $D_p$-branes) the backreaction is rather well-understood: it introduces a warp factor $e^A$ in the 10d metric
\beq
\d s^2_{10}= e^{2A} ( \d \tilde{s}^2_{4} + \d \tilde{s}^2_{6||} ) + e^{-2A} \d \tilde{s}^2_{6\bot} \ ,\label{metricwarp}
\eeq
with inverse powers along parallel and transverse directions to the sources; $\d \tilde{s}^2_{6||}$ and $\d \tilde{s}^2_{6\bot}$ still depend on all internal coordinates \cite{Andriot:2016xvq}. So we work here in this context: type II supergravities supplemented by space-time filling, parallel (i.e.~non-intersecting), backreacted $O_p/D_p$ sources, with a fixed size $3\leq p \leq 8$, and the standard dilaton value $e^{\phi}=g_s e^{A(p-3)}$ with a constant $g_s$. In this framework, we study Minkowski flux vacua; note we do not capture F-theory non-perturbative solutions. We follow conventions of a companion paper \cite{Andriot:2016xvq}.

\subsection{A. $O_3$-planes}

Let us recall known results on $O_3$ compactifications in IIB, following \cite{Giddings:2001yu, Blaback:2010sj, Andriot:2016xvq}. The sourced flux is the 5-form $F_5$, the other Ramond-Ramond (RR) fluxes being the 1- and 3-forms $F_1, F_3$, and the NSNS 3-form flux is denoted $H$. For our purposes, we need two equations: the first one is a combination of the dilaton equation of motion (e.o.m.), the 10d and the 4d traces of the Einstein equation, giving an expression for the Ricci scalar $\tilde{{\cal R}}_4$ of $\d \tilde{s}^2_{4}$; the second equation is the $F_5$ Bianchi identity (BI)
\begin{align}
& e^{-2A} \tilde{{\cal R}}_4 = -\tfrac{1}{2}|H|^2 -\tfrac{e^{2\phi}}{2}|F_3|^2 - e^{2\phi}|F_5|^2 +\tfrac{e^{\phi}}{4} T_{10}+ \dots \ , \nn\\
& \d F_5 =  H \wedge F_3 +  \tfrac{\varepsilon_3}{4}\, T_{10}\, {\rm vol}_6 \ , \label{eqO3}
\end{align}
where the dots stand for terms explicitly dependent on $\del \phi$ and $\del A$, and we leave the definitions of the square of forms, the sign $\varepsilon_p$, and the sources contributions $T_{10}$ to \cite{footnote1}. We see from above that, at least in the smeared limit where $\del \phi = \del A =0$, the flux contributions cannot be canceled to find a Minkowski vacuum without the $O_3$ contribution in $T_{10}$. We now combine both equations of \eqref{eqO3} and rewrite the result as in \cite{Andriot:2016xvq}
\bea
e^{-2A} \tilde{{\cal R}}_4   = &   - \left| e^{4A}\! *_{6}\! \d e^{-4A} - \varepsilon_3 e^{\phi}  F_{5} \right|^2 \label{finalO3}  \\
& -\tfrac{1}{2} \left|*_6 H + \varepsilon_3 e^{\phi} F_{3} \right|^2 + e^{-2A} \del(\dots) \ , \nn
\eea
where $\del(\dots)$ denotes a total derivative over the (unwarped) compact manifold. Upon integration, we deduce that a Minkowski vacuum requires both squares to vanish. One leads to the well-known ISD condition \cite{Giddings:2001yu}: $H = \varepsilon_3 e^{\phi} *_6 F_{3}$. The combination in the other square is also present in the total derivative, which thus vanishes. This also fixes $F_5$, and relates eventually $\d F_5$ to $\tilde{\Delta}_6 e^{-4A}$ as expected from the BI. With further combinations of e.o.m., one can show that $F_1=0$ and $\phi$ is constant (see below). Using the $H$ and $F_3$ BI (i.e.~that those fluxes are closed), one then shows that \emph{all e.o.m.~are solved}, provided the (unwarped) compact manifold is Ricci flat. In other words, given this geometric requirement, and assuming satisfied BI, one finds all Minkowski flux vacua, and those are characterized through the expression \eqref{finalO3} by setting the squares of BPS-like conditions to zero.

\subsection{B. $O_p$-planes with $p>3$}

In the present Letter, we derive analogous results for $O_p$-planes with $p>3$. Up to little additional structure, $\tilde{{\cal R}}_4$ can again be written in terms of squares of BPS-like conditions; with few restrictions, a Minkowski vacuum then requires to set these conditions to zero, fixing the sourced flux $F_{8-p}$, $3\leq p\leq 8$, and relating $H$ to $F_{6-p}$. We show that these three fluxes are the only non-trivial ones. Assuming their BI, we verify that \emph{all e.o.m.~are satisfied}, giving us a broad class of Minkowski flux vacua.

Let us point out the main differences when $p>3$. For $p=3$, all internal directions are transverse to the sources, while for $p>3$, some are parallel, as in \eqref{metricwarp}. $A$ is then only dependent on transverse directions, and $*_6$ gets replaced by $*_{\bot}$, $\tilde{\Delta}_6$ by $\tilde{\Delta}_{\bot}$, etc. Fluxes may now have different components: $F_l=F_l^{(0)} + F_l^{(1)} + \dots$ where $n$ in $F_l^{(n)}$ denotes the number of parallel indices of the component. We will as well introduce the projection on the transverse subspace $F_l|_{\bot}=F_l^{(0)}$. A second difference is the possibility of RR fluxes $F_l$ with $l>8-p$. Taking these features into account, a lengthy computation \cite{Andriot:2016xvq} leads to an expression for $\tilde{{\cal R}}_4$, analogous to \eqref{finalO3}, given by
\bea
e^{-2A} \tilde{{\cal R}}_4  = & - \left| e^{4A} *_{\bot} \d e^{-4A} - \varepsilon_p e^{\phi} F_{8-p}^{(0)} \right|^2 - (\mbox{flux})^2 \label{FINAL}\\
& - \tfrac{1}{2} \left|*_{\bot}H|_{\bot} + \varepsilon_p e^{\phi} F_{6-p}|_{\bot} \right|^2 + e^{-2A} \del(\dots)  \nn \\
& - \tfrac{1}{2} \sum_{a_{||}} \left| *_{\bot}( \d e^{a_{||}})|_{\bot} - \varepsilon_p e^{\phi} (\iota_{a_{||}} F_{8-p}^{(1)} ) \right|^2  \nn\\
& \ - ({\cal R}_{||} + {\cal R}_{||}^{\bot} ) + \tfrac{1}{2}|H^{(2)}|^2 + |H^{(3)}|^2    \ ,\nn
\eea
for $0\leq 8-p\leq 5$, where we use the internal one-form basis $e^a=e^a{}_m \d y^m$ constructed with Vielbeins, and the contraction by a vector $\iota_{a} e^b = \delta^b_a$. The 6d metric in flat indices is $\delta_{ab}$. One has $\d e^{a}=-\tfrac{1}{2} f^{a}{}_{bc} e^{b} \w e^{c}$, where the anholonomicity symbol (from now on ``geometric flux'') $f$ does not need to be constant. Thus by definition, $( \d e^{a_{||}})|_{\bot}=-\tfrac{1}{2} f^{a_{||}}{}_{b_{\bot}c_{\bot}} e^{b_{\bot}} \w e^{c_{\bot}}$. This expression \eqref{FINAL} is derived in \cite{Andriot:2016xvq} where on top of the analogue of \eqref{eqO3}, one uses the trace of the Einstein equation along internal parallel directions. Differences between \eqref{FINAL} and \eqref{finalO3} are related to those mentioned above: there is a third BPS-like condition involving $F_{8-p}^{(1)}$, the $(\mbox{flux})^2$ are squares of $F_{l>8-p}^{(n)}$, ${\cal R}_{||} + {\cal R}_{||}^{\bot}$ are curvature terms given in terms of $\tilde{f}^{a_{||}}{}_{b_{||}c_{||}}, \tilde{f}^{a_{||}}{}_{b_{\bot}c_{||}}, \tilde{f}^{a_{\bot}}{}_{b_{||}c_{||}}, \tilde{f}^{a_{\bot}}{}_{b_{\bot}c_{||}}$, i.e.~involving parallel directions, similarly to the components $H^{(2)}, H^{(3)}$; we refer to \cite{Andriot:2016xvq} for precise definitions.

The last line of \eqref{FINAL} is the only one with indefinite sign. We thus set it here to zero: first, we restrict to
\beq
H^{(2)} = H^{(3)} = 0 \ ,\label{Hassumption}
\eeq
while for computational convenience, we actually set the following (unwarped) geometric fluxes to zero
\bea
& \tilde{f}^{a_{||}}{}_{b_{||}c_{||}}\! =\! \tilde{f}^{a_{||}}{}_{b_{\bot}c_{||}} \! =\! \tilde{f}^{a_{\bot}}{}_{b_{||}c_{||}} \! =\! \tilde{f}^{a_{\bot}}{}_{b_{\bot}c_{||}} \! =0 \ ,\label{fassumption}\\
& \tilde{f}^{a_{\bot}}{}_{b_{\bot}c_{\bot}} \! = 0 \ , \label{fassumptionbot}
\eea
where \eqref{fassumption} implies ${\cal R}_{||}={\cal R}_{||}^{\bot}=0$; note the latter holds for all known Minkowski flux vacua on twisted tori \cite{Andriot:2015sia}. The geometric requirements \eqref{fassumption}, \eqref{fassumptionbot} can be viewed as the analogue of the Ricci flatness for $p=3$; they amount to having sources wrap a fiber, over a transverse base. The only restrictions made here on the vacua are \eqref{Hassumption}, \eqref{fassumption}, \eqref{fassumptionbot}. Integrating \eqref{FINAL} as in \cite{Andriot:2016xvq} over the 6d compact manifold then gives $\tilde{{\cal R}}_4$ only in terms of BPS-like conditions and squares of fluxes. A Minkowski vacuum requires to set them to zero, making the total derivative vanish, as for $p=3$. These conditions are enough to characterize a broad class of Minkowski flux vacua, as we now show.

\section{II. Fluxes}

Without the last line of \eqref{FINAL}, asking for a Minkowski vacuum sets the BPS-like conditions of \eqref{FINAL} and the squares of fluxes to zero. This, together with the trace of the Einstein equation along internal parallel directions \cite{Andriot:2016xvq}, implies that fluxes (and components) within the BPS-like conditions are the only ones allowed to be non-zero, i.e.~$H=H|_{\bot}$, $F_{6-p} =F_{6-p}|_{\bot}$, $F_{8-p}= F_{8-p}^{(0)} + F_{8-p}^{(1)}$, with
\bea
& F_{6-p} = - e^{-\phi} \varepsilon_p *_{\bot}H \ ,\label{flux}\\
& F_{8-p}^{(0)} = e^{-\phi} \varepsilon_p e^{4A} *_{\bot} \d e^{-4A}\ ,\nn\\
& F_{8-p}^{(1)} = e^{-\phi} \varepsilon_p\, \delta_{ab}\, e^{a_{||}} \w *_{\bot}( \d e^{b_{||}})|_{\bot}  \ .\nn
\eea
Indeed, we get that $F_{l>8-p} = 0$, and that the lowest RR flux $F_{4-p}$, meaning $F_0$ for $p=4$ and $F_1$ for $p=3$,
vanishes \cite{footnote2}. The only possible flux content is then \eqref{flux}!

The BI for these fluxes \cite{footnote2.5} are given by
\beq
\d H= \d F_{6-p}=0\ ,\ \d F_{8-p} - H\w F_{6-p} =  \tfrac{\varepsilon_p T_{10}}{p+1} \mbox{vol}_{\bot}\ ,
\eeq
and projecting the $F_{8-p}$ BI on the transverse directions, together with \eqref{flux}, gives
\bea
e^{\phi}\frac{T_{10}}{p+1} & = e^{6A} \tilde{\Delta}_{\bot} e^{-4A}  + e^{2\phi}|F_{6-p}|^2 + e^{2\phi}|F_{8-p}^{(1)}|^2 \label{BIrew}\\
& = e^{6A} \tilde{\Delta}_{\bot} e^{-4A}  + |H|^2 + |f^{a_{||}}{}_{b_{\bot}c_{\bot}}|^2 \ ,\label{BINS}
\eea
the last term being defined in \cite{footnote3}. Upon integration of the last equation one recovers the RR tadpole condition that expresses the net charge in terms of the fluxes. Both the $H$-flux and the geometric flux contribute to cancel the $O_p$ charge, and the two contributions are schematically viewed as T-dual \cite{Kachru:2002sk, Blaback:2010sj}. While $F_{8-p}^{(1)}$ is related to the geometric flux, $F_{8-p}^{(0)}$ leads to $\tilde{\Delta}_{\bot} e^{-4A}$ generating the $\delta$-functions, and can thus be interpreted as related to the sources backreaction. Explicit vacua with both pieces are given e.g. in \cite{Grana:2006kf, Andriot:2015sia}, while a vacuum with all fluxes \eqref{flux} turned on is given in (6.42) of \cite{Grana:2006kf}.

We now show that the fluxes \eqref{flux} solve their e.o.m.~\cite{Andriot:2016xvq}, given here by
\bea
& e^{-4A} \d(e^{4A} *_6 F_{8-p} )  = 0 \  (4\leq p \leq 7) \ ,\\
& e^{-4A} \d(e^{4A} *_6 F_{6-p} ) + H \w *_6 F_{8-p} = 0 \  (3\leq p \leq 5) \ ,\nn\\
& e^{-4A} \d (e^{4A-2\phi} *_6 H) -  F_{6-p} \w *_6 F_{8-p} = 0 \ ,\nn
\eea
provided relevant BI are satisfied. To that end, it is convenient to rewrite $F_{8-p}$ as follows (details in \cite{footnote4})
\beq
F_{8-p} = (-1)^p \varepsilon_p\, e^{-4A} *_6 \d\left( e^{4A -\phi} {\rm vol}_{||} \right) \ , \label{Fkcalib}
\eeq
with the internal parallel volume form, and use for an $l$-form $A_l$ that $*_6 A_l|_{\bot} = (-1)^{l(p+1)} \mbox{vol}_{||} \w *_{\bot} A_l|_{\bot}$. It is then straightforward to see that the fluxes \eqref{flux} solve the three above e.o.m., given the $H$ and $F_{6-p}$ BI. With the $F_{8-p}$ BI, we now prove the remaining e.o.m.~are also satisfied. Note it is the case for the branes own e.o.m.: \eqref{Fkcalib} is the calibration condition that minimizes their energy \cite{Andriot:2016xvq}.

\section{III. Satisfying all e.o.m.}

The remaining e.o.m.~are the dilaton and the Einstein equation; we split the latter into the 4d one, the transverse internal one, and for $p\neq3$ the internal parallel and off-diagonal ones. Verifying that they are satisfied is purely technical; we give here and in the Appendix only the main insights. For all these e.o.m., we use the 10d Einstein trace \cite{Andriot:2016xvq}, leading us to consider in the following the trace reversed Einstein equation. We also use \eqref{BIrew} and dilaton and warp factor formulas of Appendix C of \cite{Andriot:2016xvq}. The dilaton e.o.m.~and the 4d Einstein equation are then straightforward to verify. The latter boils down to
\bea
&{\cal R}|_{\mu\nu} + 2\nabla|_{\mu} \del|_{\nu}{\phi} - \tfrac{ g_{\mu\nu}}{4} (2 |\del \phi|^2 -\Delta \phi ) \label{4dEinstein} \\
& = \tfrac{g_{\mu\nu}}{8} \Big(\tfrac{e^{\phi}}{2} \tfrac{7-p}{p+1}\ T_{10} - |H|^2 + \tfrac{e^{2\phi}}{2} \sum_{q=0}^6 (1-q) |F_q|^2 \Big)\ ,\nn
\eea
with even/odd RR fluxes for IIA/IIB, and each side can be shown to be equal to \eqref{4dqtty}. We have introduced the shorthand $|_\mu \equiv {}_{M=\mu}$, and similarly $|_a \equiv {}_{A=a}$.

Internal Einstein equations are treated using flat indices. The Ricci tensor is computed using formulas of Appendix C of \cite{Andriot:2016xvq} and $f^{a_{\bot}}{}_{BC}= \delta^b_B \delta^c_C f^{a_{\bot}}{}_{bc}$. Starting from the general tensor expression, one should pay attention going from 10d to 6d indices, parallel and transverse, and further to warped versus unwarped quantities. We also use the source contribution $T_{ab} = \delta_a^{a_{||}} \delta_b^{b_{||}} \delta_{a_{||}b_{||}} T_{10}/(p+1)$. The internal parallel Einstein equation requires the dilaton derivatives \eqref{dilpar}. One should then prove that the Ricci tensor is given by \eqref{Rparpar1}, which is achieved using assumption \eqref{fassumption}. We also use \eqref{fassumption} for the internal off-diagonal Einstein equation and compute for the dilaton \eqref{diloffdiag}; this equation eventually becomes
\beq
{\cal R}|_{a_{||}b_{\bot}} = (p-2) \delta^{e_{\bot}d_{\bot}} \delta_{a_{||}c_{||}} f^{c_{||}}{}_{e_{\bot}b_{\bot}} e^{-A} \del_{d_{\bot}} e^{A} \ . \label{Einsteinoffdiag}
\eeq
The computed Ricci tensor \eqref{Rparbot} is however different. Interestingly, the match is achieved using the $F_{8-p}$ BI along non-transverse directions, i.e.~$\d F_{8-p} - (\d F_{8-p})|_{\bot} = 0$, which has two components due to $F^{(1)}_{8-p}$. Setting the one along $e^{a_{||}} \w e^{b_{||}}\w e^{\bot}$, resp. along $e^{a_{||}} \w e^{\bot}$, to zero gives identities \eqref{identity2}, resp. \eqref{identity}, using \eqref{fassumption} and $\tilde{f}^{a_{\bot}}{}_{b_{\bot}a_{\bot}} = 0$. Identity \eqref{identity} allows to solve the off-diagonal Einstein equation. Finally, using the expression \eqref{dilbot} for the dilaton, the internal transverse Einstein equation is given by
\bea
{\cal R}|_{a_{\bot} b_{\bot}} & =  - \tfrac{1}{2} \delta_{g_{||}h_{||}}  \delta^{c_{\bot}e_{\bot}}  f^{g_{||}}{}_{c_{\bot}a_{\bot}} f^{h_{||}}{}_{e_{\bot}b_{\bot}} \label{Einsteintrans}\\
& + \delta_{a_{\bot}b_{\bot}} (2p-7)  |\widetilde{\d e^{A}}|^2 + \delta_{a_{\bot}b_{\bot}} e^A \tilde{\Delta}_{\bot} e^A \nn\\
& - 2 (p-3) \del_{a_{\bot}} \del_{\tilde{b}_{\bot}} e^{A} - 2 (p+1) \del_{\tilde{a}_{\bot}} e^{A} \del_{\tilde{b}_{\bot}} e^{A} \nn\\
& +2 \omega_{a_{\bot}}{}^{c_{\bot}}{}_{b_{\bot}}|_{(\del A=0)} \del_{c_{\bot}} \phi  \nn\\
& +\tfrac{1}{2} \iota_{{a_{\bot}}}\! H \cdot \iota_{{b_{\bot}}}\! H +  \tfrac{e^{2\phi}}{2} \iota_{{a_{\bot}}}\! F_{6-p} \cdot \iota_{{b_{\bot}}}\! F_{6-p} \nn\\
& - \delta_{a_{\bot}b_{\bot}} \tfrac{e^{2\phi}}{2} |F_{6-p}|^2 \ .\nn
\eea
This holds as well for $p=3$ where $F_{5}=F_5^{(0)}$, using \eqref{flux}. The last three rows of \eqref{Einsteintrans} vanish using \eqref{fassumptionbot} and \eqref{flux}. Computing the Ricci tensor with \eqref{fassumption} and \eqref{fassumptionbot}, one then obtains a match, up to $\tfrac{1}{2}\del_{g_{||}} f^{g_{||}}{}_{a_{\bot}b_{\bot}}$, which is zero due to the identity \eqref{identity2}. All e.o.m.~are thus satisfied.

\section{IV. Non-geometric fluxes extension}

A natural extension of \eqref{FINAL} (without its last line) including non-geometric NSNS fluxes would be
\bea
2 e^{-2A} \tilde{{\cal R}}_4 & = - 2 \left| e^{4A} *_{\bot} \d e^{-4A} - \varepsilon_p e^{\phi}  F_{8-p}^{(0)} \right|^2 \label{FINALnongeo} \\
& - \left| H_{a_{\bot}b_{\bot}c_{\bot}} *_{\bot}e^{a_{\bot} b_{\bot} c_{\bot}}/3 + \varepsilon_p e^{\phi} F_{6-p}^{(0)} \right|^2 \nn\\
& - \sum_{a_{||}} \left|f^{a_{||}}{}_{b_{\bot}c_{\bot}} *_{\bot} e^{ b_{\bot} c_{\bot}}/2 + \varepsilon_p e^{\phi} (\iota_{{a_{||}}} F_{8-p}^{(1)} ) \right|^2  \nn\\
& -\! \sum_{a_{||}b_{||}}\! \left|Q_{c_{\bot}}{}^{a_{||}b_{||}} *_{\bot} e^{c_{\bot}} + \varepsilon_p e^{\phi} (\iota_{{b_{||}}} \iota_{{a_{||}}} F_{10-p}^{(2)} ) \right|^2  \nn\\
& - \!\!\! \sum_{a_{||} b_{||} c_{||}}\!\!\! \left|R^{a_{||} b_{||}c_{||}} *_{\bot} 1 + \varepsilon_p e^{\phi} (\iota_{{c_{||}}} \iota_{{b_{||}}} \iota_{{a_{||}}} F_{12-p}^{(3)} ) \right|^2  \nn\\
& - (\mbox{flux})^2  + e^{-2A} \del(\dots) \ . \nn
\eea
This combination of BPS-like conditions is T-duality invariant from the 4d perspective: there one lifts or lowers T-dualized indices of NSNS fluxes \cite{Shelton:2005cf}, while parallel and transverse directions get exchanged when T-dualized. The last line of \eqref{FINAL} may also have a T-duality invariant extension with non-geometric fluxes (see \cite{Andriot:2016xvq}). Finally, one could extend the $F_{8-p}$ BI on transverse directions \eqref{BIrew} by adding $ +  |Q_{c_{\bot}}{}^{a_{||}b_{||}}|^2  + |R^{a_{||} b_{||}c_{||}}|^2$. In these extended expressions, all terms are however not present for any $p$, due to the number of parallel and transverse directions. All fluxes are allowed for $p=6$, but not for the others: on top of the first BPS-like condition, only the one in $H$ is present for $p=3$, $H, f$ for $p=4$, $H, f, Q$ for $p=5$, $f,Q,R$ for $p=7$ and $Q,R$ for $p=8$. This restriction can also be understood schematically by T-dualizing an $O_3$ vacuum on $T^6$ with $H$-flux.

A first setting to derive these conjectured expressions would be 4d gauged supergravities, replacing $\tilde{{\cal R}}_4$ by the vacuum value of the scalar potential. Another option would be the 10d formalism of $\beta$-supergravity \cite{Andriot:2013xca}, or extensions thereof, including \cite{Geissbuhler:2013uka}. But this goes beyond the scope of this Letter.

\section{V. Discussion}

In this Letter, we have characterized a broad class of Minkowski flux vacua with parallel localized $O_p/D_p$ sources; this is achieved thanks to the general rewriting \eqref{FINAL} of the 4d Ricci scalar as a negative sum of squares, extending a well-known result for $O_3$-planes \cite{Giddings:2001yu} to general $O_p$ reductions. The fluxes are \eqref{flux} (their BI are assumed to be satisfied), the metric is \eqref{metricwarp} with restrictions \eqref{fassumption} and \eqref{fassumptionbot}, and the dilaton is given after \eqref{metricwarp}. Interestingly, requiring only the restrictions \eqref{Hassumption}, \eqref{fassumption} and \eqref{fassumptionbot}, we have shown that these are {\em the only Minkowski flux vacua}!

Our analysis does not rely on supersymmetry (SUSY) so our vacua can capture both SUSY and non-SUSY solutions. To get SUSY $O_3$ vacua, the Ricci flat condition should be supplemented by Calabi-Yau (see \cite{Andriot:2015sia} for a Ricci flat non-Calabi-Yau solvmanifold), and the ISD 3-form flux has to be (1,2) and primitive \cite{Giddings:2001yu}. Here, it would be interesting to compare our geometric restrictions to the generalized Calabi-Yau condition and our fluxes to the SUSY ones \cite{Grana:2006kf, Andriot:2015sia}, given the integrability result of \cite{Koerber:2007hd}. Already, our sourced flux takes the value \eqref{Fkcalib} of a calibrated brane \cite{Andriot:2016xvq}, which would be the SUSY value in presence of an SU(3)$\times$SU(3) structure.

A natural question is whether there exists other Minkowski flux vacua with parallel sources than the class found here, thus meaning some with the restrictions not satisfied (see a related discussion in \cite{Andriot:2016xvq}). We may capture all no-scale vacua, but still wonder whether other vacua exist, which do not arise from the BPS-like conditions. We can for instance look at the set of SUSY Minkowski flux vacua on twisted tori: all known solutions are reviewed in \cite{Andriot:2015sia}. All listed vacua with parallel sources verify \eqref{Hassumption}, \eqref{fassumption}, \eqref{fassumptionbot} (in particular those of \cite{Grana:2006kf}), except for the new ones found in \cite{Andriot:2015sia} that violate \eqref{fassumption}. For the latter however, the Ricci tensor still vanishes and ${\cal R}_{||}={\cal R}_{||}^{\bot}=0$, indicating a suspected possible refinement of our geometric conditions. Then, up to this detail, we do not find vacua beyond the class presented here.

\section{Acknowledgements}

We thank F.~F.~Gautason and D.~Tsimpis for useful discussions.  The work of D.~A.~is part of the Einstein Research Project ``Gravitation and High Energy Physics'', funded by the Einstein Foundation Berlin. The work of J.~B.~was supported by the John Templeton Foundation Grant 48222 and the CEA Eurotalents program. The work of T.~V.~R.~is supported by the FWO odysseus grant G.0.E52.14N. We acknowledge support from the European Science Foundation Holograv Network.

\appendix*
\setcounter{equation}{0}
\section{Appendix: Ricci tensor and dilaton derivatives}

Each side of the 4d Einstein equation \eqref{4dEinstein} is equal to
\beq
\tfrac{g_{\mu\nu}}{16} (7-p) \left( e^{6A} \tilde{\Delta}_{\bot} e^{-4A} - e^{8A}|\d e^{-4A}|^2 \right) \ .\label{4dqtty}
\eeq
For the internal parallel Einstein equation, we use
\bea
& \nabla|_{a_{||}}\del|_{b_{||}} \phi = - \tfrac{\delta_{a_{||} b_{||}}}{4} e^{2\phi-2A} \delta^{c_{\bot}d_{\bot}} \del_{c_{\bot}}\! e^{2A} \del_{d_{\bot}}\! e^{-2\phi}\label{dilpar}\\
& {\cal R}|_{a_{||} b_{||}} = \tfrac{ \delta_{a_{||} b_{||}}}{8} \left( 2 e^{6A} \tilde{\Delta}_{\bot} e^{-4A} - (p-1) e^{8A}|\d e^{-4A}|^2 \right) \nn\\
& \phantom{{\cal R}|_{a_{||} b_{||}}}\!\! + \tfrac{\delta_{a_{||} c_{||}} \delta_{b_{||} d_{||}} }{4} \delta^{e_{\bot}f_{\bot}} \delta^{g_{\bot}h_{\bot}} f^{c_{||}}{}_{e_{\bot}g_{\bot}} f^{d_{||}}{}_{f_{\bot}h_{\bot}} \ ,\label{Rparpar1}
\eea
where the warp factor terms of \eqref{Rparpar1} can be rewritten as $- \delta_{a_{||} b_{||}} \delta^{c_{\bot} d_{\bot}} \left( e^{A} \del_{\tilde{c}_{\bot}} \del_{\tilde{d}_{\bot}} e^{A} + (2p-7) \del_{\tilde{c}_{\bot}}  e^{A} \del_{\tilde{d}_{\bot}} e^A \right)$. For the internal off-diagonal Einstein equation, we compute with \eqref{fassumption} and $2\del_{[a} \del_{b]}= f^c{}_{ab}\del_c$
\bea
& \hspace{-0.1in} \nabla|_{a_{||}} \del|_{b_{\bot}}\! \phi = \! \nabla|_{b_{\bot}} \! \del|_{a_{||}} \phi = -\tfrac{\delta_{a_{||}e_{||}}\delta^{c_{\bot}d_{\bot}}}{2} \! f^{e_{||}}{}_{d_{\bot}b_{\bot}}\! \del_{c_{\bot}}\! \phi  \label{diloffdiag}\\
& \hspace{-0.1in} {\cal R}|_{c_{||}d_{\bot}} = (p-3) \delta^{b_{\bot}e_{\bot}} \delta_{c_{||}h_{||}} f^{h_{||}}{}_{e_{\bot}d_{\bot}} \del_{\tilde{b}_{\bot}} e^{A} \label{Rparbot}\\
& \hspace{-0.1in} \phantom{{\cal R}|_{c_{||}d_{\bot}}} + \tfrac{1}{2} \delta^{b_{\bot}e_{\bot}} \delta_{c_{||}h_{||}} \del_{b_{\bot}} f^{h_{||}}{}_{e_{\bot}d_{\bot}} \nn\\
& \hspace{-0.1in} \phantom{{\cal R}|_{c_{||}d_{\bot}}} + \tfrac{1}{4} \delta_{c_{||}i_{||}} \delta_{d_{\bot}g_{\bot}} \delta^{e_{\bot}h_{\bot}} \delta^{b_{\bot}j_{\bot}} f^{i_{||}}{}_{e_{\bot}j_{\bot}} f^{g_{\bot}}{}_{h_{\bot}b_{\bot}} \ ,\nn
\eea
and the $F_{8-p}$ BI along non-transverse directions gives
\bea
& \del_{a_{||}} f^{c_{||}}{}_{d_{\bot}e_{\bot}} = 0 \ , \label{identity2} \\
- & 2 \delta^{l_{\bot}d_{\bot}} f^{b_{||}}{}_{c_{\bot}l_{\bot}} \del_{\tilde{d}_{\bot}} e^{A} + \delta^{d_{\bot}l_{\bot}} \del_{d_{\bot}} f^{b_{||}}{}_{c_{\bot}l_{\bot}} \label{identity}\\
- & \tfrac{1}{2} \delta_{k_{\bot}c_{\bot}} \delta^{e_{\bot}h_{\bot}} \delta^{f_{\bot}g_{\bot}} f^{b_{||}}{}_{h_{\bot}g_{\bot}} f^{k_{\bot}}{}_{e_{\bot}f_{\bot}} = 0 \ . \nn
\eea
For the transverse Einstein equation, we compute
\bea
\nabla|_{a_{\bot}} \del|_{b_{\bot}} \phi &= \del_{a_{\bot}} \del_{b_{\bot}} \phi - \omega_{a_{\bot}}{}^{c_{\bot}}{}_{b_{\bot}}|_{(\del A=0)} \del_{c_{\bot}} \phi \label{dilbot}\\
& + \del_{a_{\bot} } \phi \del_{\tilde{b}_{\bot}} e^A- \delta_{a_{\bot} b_{\bot}} \delta^{c_{\bot} d_{\bot}} \del_{c_{\bot} } \phi \del_{\tilde{d}_{\bot}} e^A \ .\nn
\eea


\begin{thebibliography}{100}
\bibitem{footnote1}
We define the square of the $q$-form $A_q$ as $|A_q|^2=A_{m_1\ldots m_q}A^{m_1\ldots m_p}/q!$, the sign $\varepsilon_p=(-1)^{\left[\frac{9-p}{2} \right]+p+1}$, and sources are localized in transverse directions through $T_{10} = \tfrac{2 \kappa_{10}^2}{\sqrt{|g_{\bot}|}}\, (p+1)  T_p  \Big(\sum_{O_p} \! 2^{p-5} \delta (\bot) - \sum_{D_p} \! \delta (\bot) \Big)$. We consider BPS sources where tension and charge are related as $T_p=\mu_p$, and without world-volume flux.
\bibitem{footnote2.5}
The non-zero fluxes may also enter the $F_{10-p}$ BI through $H\w F_{8-p}$, but this quantity vanishes as it should, given its number of transverse components.
\end{thebibliography}
\end{document}